\input{psfig}

\documentstyle[twoside,fleqn]{article}
\def\FIGREF#1{\ref{#1}}

\newenvironment{equl}[1]{\begin{equation}\label{#1}}{\end{equation}}
\def\CITE#1{{\cite{#1}}}
\def\mxnote#1{}
\def\mxnote#1{\marginpar{\tiny ~~#1 }}
\def\REF#1{(\ref{#1})}
\def\FAZA#1#2{{ie\over \hbar c}\int_{#1}^{#2}{\bf A \cdot dl } }
\def\xnote#1{}

\newcommand{\nn}{{\vec{n}}}

\def\srv#1{\langle #1 \rangle}

\def\sT0#1{\langle T_\tau [#1] \rangle_0}
\def\kr{^\dagger}
\def\ide{\rightarrow}

 \def\FAZA#1#2{{ie\over \hbar c}\int_{#1}^{#2}{\bf A \cdot dl } }

\newcommand{\AmS}{{\protect\the\textfont2
  A\kern-.1667em\lower.5ex\hbox{M}\kern-.125emS}}

\begin{document}

{\Large
\centerline{ Raman spectrum and charge fluctuations in the}
\centerline{ copper-oxide superconductors}
}

\leftline{H.~Nik\v si\' c$^{\rm a}$, E.~Tuti\v s$^{\rm b}$
        and S.~Bari\v si\' c$^{\rm a}$
}

\vskip 0.5cm

\leftline{$^{\rm a}$Department of Physics, Faculty of Science,
        Bijeni\v cka 32, POB 162, Zagreb, Croatia
}

\leftline{$^{\rm b}$Institute of Physics,
        Bijeni\v cka 46, POB 304, Zagreb, Croatia
}

\vskip 0.5cm

{
\centerline{ABSTRACT}

The effect of the charge fluctuations on
the electronic spectrum and the Raman
spectrum of high temperature
superconductors is examined within the
slave boson approach. Instead of using the saddle
point approximation~for slave bosons,
we confine ourselves to
the non-crossing
approximation (NCA) in summing the diagrams for the Green
functions, thus obtaining the renormalized
hole spectrum and its~lifetime on equal footing.
The electronic Raman spectrum is calculated,
showing the characteristic featureless behaviour
up to the frequency of the order of renormalized $\Delta_{pd}$ parameter.
The dependence~on the polarization of the incident
and the scattered light agrees with experiments.



\section{Introduction}

    One of the interesting properties of the $HTSC$ copper-oxide
materials is the featureless Raman spectrum
for frequencies up to 1 $eV$, with characteristic
polarization dependences
\CITE{SHM2,COP}.
      Here we reproduce qualitatively such behaviour within
the slave boson formalism \CITE{ReadNewns} for the {\it p-d} model
with large $U_d$.
    In treating the slave bosons,
we choose non-crossing ($NCA$) diagrams for Green functions,
following Kroha et al. \CITE{Kroha92}
approach to the Anderson model.
    The resulting spectrum for the electron Green functions has
three bands, ``p''-band, ``d''-band and the ``in-gap''-band
which contains the Fermi-level,
consistently with some previous theoretical results
and experimental observations.

\section{Model and method}

   Within the {\it p-d} model for electrons in $CuO_2$
we retain \CITE{KLR88}
the Coulomb repulsion $U_d$,
the parameters $\epsilon_d$, $\epsilon_p$
and $t_0$, denoting respectively
the energies of the hole in  $Cu$ $3d_{x^2-y^2}$
and $O$ $2p(\sigma)$ orbitals
and the copper-oxygen hybridization.
     Taking the $U_d\ide \infty$ limit,
the slave boson substitution \CITE{ReadNewns}
$d_{\nn\sigma}\ide b\kr_\nn f_{\nn\sigma}$
is accompanied by the appearance~ ~of~ ~an~ ~additional,~
{}~static,~ ~$\lambda$-field \\
($H\ide H+ \sum_{\nn}\lambda_\nn \hat{Q}_\nn$)
providing for the physical constraint
$\hat{Q}_\nn\equiv\sum_{\sigma}
f_{\nn\sigma}\kr f_{\nn\sigma} + b_\nn\kr b_\nn=1$.
     Following Kroha et al. \CITE{Kroha92}, we neglect the
fluctuations of $\lambda_\nn$ around the mean field value $\Lambda$.
This means that  the states with
$Q_\nn$ different from unity became allowed.
     However, with $\lambda_\nn$ fixed to $\Lambda$,
the Hamiltonian, \hbox{$H=H_0+H_{hyb}$}
(with $H_{hyb}$ describing the hybridization
of the copper orbital with the
{\it bonding combination} of the oxygen orbitals),
\begin{eqnarray}
\label{H0}
 H_0  \hbox to -0.7mm{}    = \hbox to -2.5mm{}
 \displaystyle \sum_{k,\sigma}\hbox to -0.7mm{} \{ \hbox to -0.3mm{}
 \epsilon_p \hbox to -0.1mm{} p_{k,\sigma}\kr p_{k,\sigma} \hbox to -0.7mm{}
 + \hbox to -0.6mm{} ( \hbox to -0.4mm{} \epsilon_d \hbox to -0.7mm{}
  + \hbox to -0.9mm{} \Lambda \hbox to -0.3mm{} )
 \hbox to -0.3mm{} f_{k,\sigma}\kr f_{k,\sigma} \hbox to -0.6mm{}
 \} \hbox to -0.7mm{} +\hbox to -2.6mm{} \sum_{k} \hbox to -0.3mm{}
 \Lambda \hbox to -0.0mm{} b_k\kr b_k \hbox to -1.0mm{} \\
\label{Hhyb}
 H_{hyb}  =  \displaystyle \sum_{k,\sigma} [ t(k)  p_{k,\sigma}\kr
 \sum_{q}f_{k+q,\sigma}b_{q}\kr + h.c. ]~~~~~~~~  \\
 t(k)\equiv 2t_0\sqrt{\sin(k_x/2)^2+\sin(k_y/2)^2}~~~~~~~~~~~~
\end{eqnarray}
conserves $\hat{Q}_\nn$, so the states
with different $Q_\nn$ are not dynamically mixed.
  Therefore, we expect that on fixing
$\srv{\hat{Q}_\nn}=1$ our calculation will predominantly
reflect the dynamics of the $Q_\nn=1$ system.
  Further, following Kroha et al.  we  confine
ourselves  to $NCA$ diagrams,
thus discarding the {\it Cu-Cu} spin flip processes.
    Within the $NCA$ we get the set of Dyson equations (Fig. \FIGREF{NCAfig}),
               \begin{figure}[htb]
			\psfig{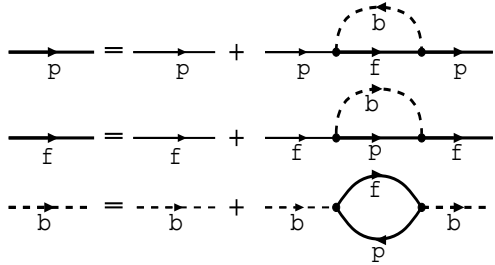}
               \caption{$NCA$ diagrams for the Green functions. \label{NCAfig}}
               \end{figure}
which~ we~ solve~ numerically.~
    In~ absence~ of~ the~ \\
$b$-condensate,
the one-particle Green functions for $b$
and $f$ fields are site-diagonal
to all orders in the perturbation in Eq.~\REF{Hhyb}.
     The Green function for the hole on the copper-site
is obtained from the bare Green function $G_p^0$ on the oxygen-site
and its self-energy corrections $\Sigma_p$ from Fig. \FIGREF{NCAfig}.
\begin{equl}{Gd}
G_d(\omega,k)^{-1}=
\Sigma_p(\omega,k)^{-1}-\mid t(k)\mid^2G_p^0(\omega,k)~.
\end{equl}

\section{Hole spectrum and density of states}

   The~density~of~states~($D.O.S.$)~for~the~copper and the
oxygen site which we obtain for $1+\delta=1.14$
holes per unit cell  are shown in
Fig.\FIGREF{dosGdGcspfig}.
     We~ ~choose~ ~parameters~
{}~$\Delta_{pd}=\epsilon_p-\epsilon_d=4.0t_0$ \\
($t_0\approx 1eV$), $T=0.0135t_0$ (temperature)
and set $\epsilon_p=0$ for convenience.
     The values for $\Lambda$
and $\mu$ (Fermi-level) are $\Lambda=3.75t_0$ and $\mu=-1.138t_0$.
	For obvious physical reasons the $D.O.S.$
accumulates~ ~around~ ~energies~ ~$\epsilon_p$~
{}~({\it ``p''-band}),~
{}~and~ ~$\epsilon_d$ \\
({\it ``d''-band}) of the unhybridized
oxygen and copper orbitals.
               \begin{figure}[htb]
			\psfig{figure=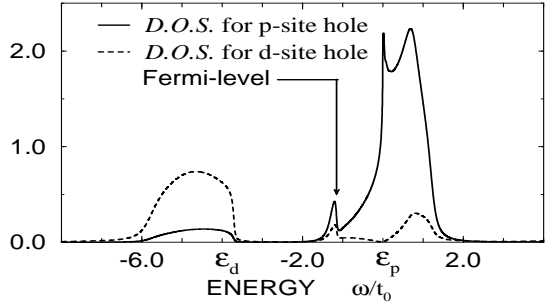,width=7cm,height=4.5cm}
               \caption{$D.O.S.$ for $p$-site and $d$-site holes.
			\label{dosGdGcspfig}}
               \end{figure}
	The Fermi level lies in a separate,
resonant {\it ``in-gap''-band},
similar to the result of the
saddle point calculations in {\it b} \CITE{KLR88}.
Note, however, that the band edges are smoothened out,
due to the ``disorder'' produced by the slave boson.

\section{Raman spectrum}

The coupling of  the electromagnetic field
to electrons is introduced
by the substitution,
\begin{equl}{addA}
tc_r^{\dagger}c_s\ide tc_r^{\dagger}c_se^{\FAZA{s}{r} }~.
\end{equl}
     Expanding up to the quadratic terms in vector potential,
we get usual ~$\vec{p}\cdot\vec{A}$~
and ~$n{\vec{A}}^2$~ electron-light coupling terms.
     Using these we calculate the resonant
and nonresonant contributions to the transition matrix
for the Raman scattering.
     The overall results are shown in Fig. \FIGREF{Rampolfig}.
               \begin{figure}[htb]

\centerline{\psfig{figure=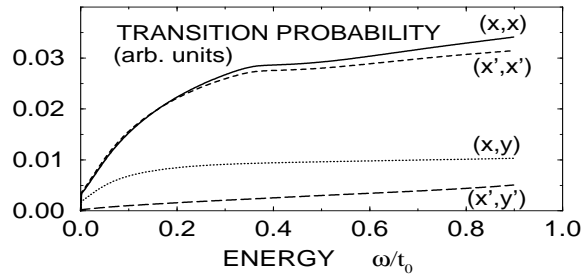,width=7cm,height=5cm}}
               \caption{Transition probability for the Raman scattering for
various polarizations.
			\label{Rampolfig}}
               \end{figure}
       The featureless Raman spectrum,
in general agreement with experiment,
extends up to the frequency of $1~eV$
approximately.
    In our calculations this scale corresponds
to the separation of the ``in-gap''-band and ``p''-band
(corresponding to renormalized $\Delta_{pd}$
in the saddle point approximation \CITE{KLR88}).
     Above that frequency (not probed experimentally)
a steep increase of the Raman intensity
occurs in our calculations
due to the interband processes.
  The intensity of the spectrum
significantly  changes  as a function of
the polarization of  the incident and the
scattered light.  This is  due to the
polarization dependence of the
electron-light  interaction,
characteristic for the nearest neighbour
hoping Hamiltonian in Eq.~\REF{Hhyb}
(for example, the scattering via the
$n \vec{A}^2$ coupling does not
contribute at all for the (x',y') polarization).
     The polarization dependence in our calculations
(shown above) is in a qualitative accordinance
with experiments \CITE{SHM2,COP}.

\def\RCITE#1{\bibitem{#1}}

\end{document}